\documentstyle[12pt,epsf]{article}
 \newcommand{\be}{\begin{equation}}
 \newcommand{\ee}{\end{equation}}
 \title{Earthquake statistics and fractal faults}
 \author{R. Hallgass$^1$, V. Loreto$^{2,3}$, O. Mazzella$^4$, G. Paladin$^1$ 
 and L. Pietronero$^3$}
 
\begin{document}
 \maketitle
 \centerline{$^1$Dipartimento di Fisica,  Universit\`a dell'Aquila}
 \centerline{Via Vetoio I-67100 Coppito, L'Aquila, Italy}
 \centerline{$^2$ Centro Ricerche  ENEA, Localit\`a Granatello}
 \centerline{C.P. 32 - 80055 Portici (Napoli) Italy}
 \centerline{$^3$Dipartimento di Fisica, Universit\`a di Roma 
 'La Sapienza'}
 \centerline{P.le A.Moro 2 I-00185 Roma, Italy}
 \centerline{$^4$ Dipartimento di Fisica, Universit\`a di Como}
 \centerline{ Como, Italy}
 \date{}
 \medskip
 
 \begin{abstract}
 We introduce a Self-affine Asperity Model (SAM) for the seismicity
that mimics the fault friction by means of
two fractional Brownian profiles (fBm) that slide one
 over the other. An earthquake occurs when there is
 an overlap of the two profiles representing the two fault faces
and its energy is assumed proportional to the  overlap surface.

The SAM exhibits the Gutenberg-Richter law with an exponent
$\beta$ related  to the roughness index of
 the profiles. Apart from being analytically treatable, the
 model exhibits
a non-trivial clustering in the spatio-temporal distribution 
of epicenters that strongly resembles the experimentally
 observed one.
 A generalized and more realistic version of the
model exhibits the Omori scaling for the distribution of the 
aftershocks.

The SAM lies in a different perspective with respect to usual models
for seismicity. In this case, in fact, the critical behaviour is not 
Self-Organized but stems from the fractal geometry of the faults,
 which, on its turn, is supposed to arise as a consequence of 
geological processes on very long time scales with respect to the
 seismic dynamics. The explicit introduction of the fault geometry,
as an active element of this complex phenomenology, represents the
real novelty of our approach.

\end{abstract}
 \vskip .4truecm
 \noindent
 PACS NUMBERS: 91.39.Px, 05.45.+j

\newpage

 \section{Introduction}

Recently, theoretical models have acquired an increasing
 relevance  in the study of the seismicity.
Their aim is to fill up the gap between the experimental
knowledge and the theoretical comprehension of the
phenomenon.
 
 One of the most serious problem which geologists have to face
is the lack of complete catalogues extended over  long time 
periods. This makes difficult to improve the general comprehension
 about earthquakes. By studying theoretical models, one then tries
to focus on some particular ingredients, which are supposed to be
essentials, and then tries to understand as much as possible of the
 seismic behaviour. In this way one can compare the specific
predictions of the models with those obtained from
the real catalogues.

Though the dynamics of earthquakes is very complex there are 
some simple basic components which have to be taken into 
account in a model: 

\begin{itemize}

\item[(a)] earthquakes are generated by a very slow 
discontinuous driving of 
a fault;
\item[(b)] the occurrence of earthquakes is intermittent, i.e. 
they occur
as abrupt rupture events when the fault can no longer sustain 
the stress;
\item[(c)] there are two separate time scales involved in the 
process; one
is related to the stress accumulation while the other, which is 
orders of 
magnitude smaller, is associated to the duration of the abrupt
 releases
of stress.
\end{itemize}

Many forms of scaling invariance appear in seismic phenomena.
The  most impressive feature is the celebrated 
Gutenberg-Richter 
law \cite{GR} for the magnitude distribution of earthquakes. It 
states  
that the probability $P(E)dE$ that an earthquake releases an 
energy in 
the interval $\left[ E, E+dE \right]$ scales according to a 
power-law
\be
P(E) \sim E^{-\beta-1}
\ee
with an exponent $\beta$ of order of unity  
whose eventual universality represents matter of debate.

The Omori law \cite{omori} for the time correlations of 
aftershocks 
(i.e. seismic events which happen as a consequence of a main
 earthquake)
is another example of scaling behaviour in the seismic 
phenomenology and 
one of the most difficult to reproduce in simplified models.

In the last decades there has been an increasing evidence for 
 the space-time clustering \cite{cluster}  of the earthquake
 epicenters. 
In particular there are experimental evidences suggesting 
 that the epicenter distributions is self-similar both in space 
and in time.

Unfortunately, the complexity  of  modelling the motion of a 
fault 
system, even in rather well controlled situations such as the San 
Andreas fault in California, is a highly difficult task and it is 
still controversial what is the correct theoretical framework at 
the very origin of scaling laws. It is thus important to 
individuate models as simple as possible that are able to
 exhibit 
the main  qualitative features of the fault dynamics.
Their physical relevance stems from the specific predictions 
on the {\it real} seismic activity which might be verified 
from experimental data. 
 
One of the first attempt in this direction is due to Burridge and 
Knopoff \cite{burkno} who introduced a stick-slip model  of
 coupled 
oscillators to mimic the interaction of two fault surfaces. 
In  practice, 
one considers blocks on a rough support connected to one 
 another 
by springs. They are also connected by other springs to a driver 
which moves at a very low constant speed. The blocks stick 
until 
the spring forces overwhelms the static friction and then one or 
more 
blocks slide, releasing an `earthquake' energy proportional to
 the 
sum of the displacements. In the frame of the inferior plate,
if $x_i$ denotes the position of th $i$-th block, the 
equations of motion are
\be
m_i \ddot {x_i} = k_{c,i} (x_{i+1}- x_{i}) -k_{c, i-1} (x_i -x_{i-1})
-k_{p,i} (x_i -vt)+ F_i (\dot{x_i})
\ee
where $F_i (\dot{x_i})$ represents the friction force which
 depends
on the block velocity $\dot{x_i}$. In the original model the friction
force, zero for zero velocity, increases progressively as the
 velocity
increases up to a certain maximum value. This model exhibits
 the
Gutenberg-Richter law for the distribution of the energy
 released
during an earthquake and it allows for the presence of 
aftershocks.
Up to now the original model of Burridge and Knopoff remains 
the only one
able to explain the presence of aftershocks without  
{\em ad hoc} modifications.

A numerical integration of the Newton 
equations for a one-dimensional  chain with a large number of 
homogeneous blocks has been performed by Carlson and 
Langer
\cite{carlson}. Their model differs from that
of Burridge and Knopoff in the form of the friction force which
 is 
supposed to be identical for all the blocks, neglecting the
inhomogeneities of the crust. It has been shown that the model
exhibits the Gutenberg-Richter law \cite{GR} 
(see also \cite{paladin} for the connection with the 
chaotic behaviour of the system). 

More recently it has been suggested that the qualitative aspects
 of
earthquakes (and of Burridge and Knopoff models) could be
 captured by the 
so-called Sandpile models, which are the paradigm of a
 large 
class of models showing Self-Organized Critical (SOC) \cite{btw}. 
The concept of Self-Organized Criticality (SOC) has been invoked 
by Bak, Tang and Wiesenfeld \cite{btw}  to describe the tendency
of dynamically driven systems to evolve spontaneously towards a critical 
stationary state with no characteristic time or length scale. 
An example of this behaviour is provided by Sandpile models: sand is
added grain by grain in a pile on a $d$-dimensional lattice until unstable
sand (too large local slope of the pile) slides off.
In this way the pile reaches a steady state where additional sand 
grains fall off the pile by avalanche events. This steady state is 
critical since avalanches of any size are observed. According to this 
picture of Self-Organized Criticality, during its whole evolution the 
earth would have reached a marginally stable state in which any small 
perturbation could give rise to relaxation processes, earthquakes in 
this case, that can be small or cover the entire system. In this 
way the earthquakes would be the equivalent of  avalanches 
for Sandpiles models. The main ingredient in this picture would be 
the interplay between the slow dynamics, represented by the stress
accumulation, and the fast dynamics of earthquakes.
The latter would modify the earth crust which, on its turn, can give rise to 
earthquakes and so on, with a feedback mechanism that would be at the
origin of the self-organization.

There exists a whole generation of SOC models proposed to explain the 
scale-invariant properties of earthquakes \cite{baktang,ito}. 
These type of models suggest however that there is no stress accumulation 
before a big earthquake and the exponent of the Gutenberg-Richter law is 
expected (with the exception \cite{olami} that we mention hereafter)
 to be universal. 
In addition the space-time distribution of the epicenters has no clear 
relation with the experiments where non-trivial clustering are present.

It is worth to recall in this  framework the model  
 proposed by Olami, Feder and Christensen \cite{olami}.
Their model maps the two dimensional version of the Burridge-Knopoff
spring-block model in a cellular automaton and it gives a good 
prediction of the Gutenberg-Richter law with a non-universal
value of the $\beta$-exponent, which varies with the level of 
non-conservation of the model and could account for the $\beta$ variances
observed in nature.

In order to go beyond the limitations of these models,
we have recently proposed an alternative approach \cite{prl} where the
critical behaviour is not self-organized but stems from the fractal
geometry of the faults \cite{brown, power,huangturcotte}. 
In this perspective the faults are supposed to be
formed as a consequence of geological processes on very long time 
scales  with respect to the seismic dynamics.
Looking at the system on the time scale of human records the fault 
structure can be considered assigned and just slightly modified by
earthquakes.
 
In particular, we have introduced the so-called Self-affine Asperity Model
(SAM) \cite{prl} which mimics the fault dynamics by means of the slipping
of two rough and rigid brownian profiles one over the other.
In this scheme an earthquake occurs when there is an intersection between
the two profiles. The energy released is proportional to the overlap 
interval. This model, apart from being analytically treatable, 
exhibits some specific features which follow from the fractal
geometry of the fault. In particular it reproduces the Gutenberg-Richter
law with an exponent $\beta$ which is non-universal since it depends
 on the roughness of the fault profiles. It predicts the presence of a
 local stress accumulation before a large seismic event. Moreover it 
allows one to analyze and investigate the complex phenomenology of 
the space-time clustering of epicenters. 
The model exhibits, in fact, a long-range correlation
of the events which corresponds to a self-similar distribution of the spatial
and temporal epicenter sets.
In this scheme it is also possible to include the analysis of the origin of 
aftershocks and show that, in a natural generalization of the model,
 they follow the celebrated Omori's law.

In this paper we describe in detail the SAM.
The analytical results are, step by step, tested numerically
and, whenever possible via  the comparison with experimental data.

The outline of the paper is the following. 
In sect. II we introduce the model and we recall some properties of
fractional Brownian profiles. Sect. III is devoted to the discussion
of the Gutenberg-Richter law. We show that the SAM follows 
this scaling with an exponent $\beta$ that we relate
 analytically to the roughness of the brownian profile. This 
allows us to draw some conclusions on the non-universality of 
the exponent
$\beta$. In sect. IV we discuss the problem of the distribution 
of epicenters both from the spatial point of view and the 
temporal one. The SAM exhibits a non-trivial 
clustering of epicenters which reproduces the experimental 
results and can be analytically explained by exploiting the 
properties of the fractional Brownian profiles. The problem of 
the power spectrum of the temporal sequence of earthquakes is
 also discussed. Sect. V is dedicated to the introduction of a
 more realistic version of the SAM model. This version, which 
takes into account  the local rearrangement of the earth crust 
as a consequence of the earthquakes, exhibits a non-trivial 
scaling in the distribution of the aftershocks, according 
the  Omori's law. Finally in sect. VI we draw 
the conclusions. The paper is completed by two appendices on the 
statistics of the fractional Brownian profiles. 
\newpage
 
\section{The model}

  Many authors pointed out that natural rock surfaces 
  can be represented by fractional brownian surfaces over a
 wide scale range \cite{brown,power} and that also the 
topographic traces of the fault surfaces exhibit scale invariance 
\cite{mandel}.
  A fault can thus be regarded as a 
  statistically self-affine profile $F_H (t)$, whose height scales as 
  $|F_H(t+\tau)-F_H(t)|\sim \tau^{H}$. In $d=2$, such a profile 
  $F_H(t)$ can be generated by fractional Brownian motion (fBm) 
with exponent $H$, the Hurst exponent, and in $d=3$ by the standard
 generalization given by brownian reliefs \cite{turcotte,smalley}.
  The exponent $0\le H \le 1$ controls the roughness of the fault 
  where the standard brownian profile corresponds to $H=1/2$, 
  and a differentiable curve corresponds to $H=1$.
Just to give an example let us recall how it is possible to generate
a brownian profile. In the one-dimensional case one can generate 
$L$ random variables (r.v.) $\lbrace X_1,...,X_L \rbrace$ according
to the following algorithm:
\begin{displaymath}
X_i = \left\{ 
\begin{array}{ll}
1 &  \mbox{with prob. $p=1/3$} \\
0 &  \mbox{with prob. $p=1/3$} \\
-1 &  \mbox{with prob. $p=1/3$} 
\end{array}
\right.
\end{displaymath}
On a one-dimensional lattice of $L$ sites one can thus define a 
stochastic function
\be
S(n) = S_0 + \sum_{i=1}^{n} X_i \,\,\, \forall n \le L
\label{S}
\ee
where $S_0$ is an arbitrary integer number. The (\ref{S}) defines,
in the limit $n \rightarrow \infty$ a self-affine profile of fractal
 dimension $D=1.5$. More in general
the fractal dimension of the profile is well known to be $D_F=d-H$. 
For further details we refer to the appendix A.

The explicit introduction of the fault geometry in a model for
seismicity was already been supposed by Huang and Turcotte
 \cite{brown}. They introduced
   a static model where the average of all the seismic events 
   contributing to the Gutenberg-Richter law is taken over many 
   uncorrelated realizations of one single fractal profile.
    The purpose of this letter is to introduce a dynamical model,
    called Self-affine Asperity Model (SAM), that describes 
   the seismic activity considering two profiles 
   sliding one over the other instead of only one as in \cite{brown}. 
   Such a model has the advantage to exhibit strong spatial and temporal
   correlations also between far away seismic events, and allows
    us to infer some specific and new predictions about the relation
    between  the roughness of the  fault $H$  
    and the scaling exponent of the Gutenberg-Richter law as well as on 
    the spatio-temporal distribution of epicenters.

Note how this model represents an alternative approach with respect
to the SOC models. In this case, in fact, one supposes as lacking the 
interplay between the fault structure and the seismic events. 
The latter is supposed not to modify substantially the fault geometry.
In this sense one is in a sort of  limit of infinite rigidity 
  of the Burridge-Knopoff models.
 
  Operatively, the  SAM is defined by the following dynamical rules:
  
{\bf (i)} We consider two profiles, say $S^{\prime} (n)$ and
 $S^{\prime \prime}(n)$, with $n=1,...,L$,  on parallel supports 
of length $L$ at infinite distance. 
  The initial condition is obtained by putting them in contact in 
  the point where the height difference is minimal so that (see Fig.1)
\begin{displaymath}
\left\{
\begin{array}{l}
S^{\prime} (n)=  S^{\prime \prime } (n) + 
max_{j \in \lbrace 1,...,L \rbrace } \lbrace S^{\prime} (j) - S^{\prime
 \prime} (j) \rbrace \\
n=1,...,L
\end{array}
\right.
\label{pippo}
\end{displaymath}

  {\bf (ii)} The successive evolution is obtained by drifting 
  a profile in a parallel way with respect to the other one, 
  at a constant speed $v$, so that  $S^{\prime}(n;t)=S^{\prime}(n-v \,t)$;

  {\bf (iii)} At each time step $t$, one controls whether there
  are new contact points between the profiles, i.e.
  whether $S^{\prime}(n;t)-S^{\prime \prime}(n) < 0$ for some 
$x$ value.
  An intersection represents a single seismic event
  and starts with the collision of two {\it asperities} of the profiles. 
  The energy released is assumed to be proportional to 
  the breaking area of the asperities, 
  i.e. the extension of the hypersurfaces, 
  in general of dimension $(d-1)$, involved in the collision of 
  the asperities during an earthquake. 
  In the case $d=2$ the energy released is given by the sum of the
  lengths of the two segments indicated with $A$ and $B$ in Fig.1;

  {\bf (iv)} We do not allow the developing of new earthquakes 
  in a region where a seismic event is already taking place, i.e.,
  with reference to the Fig.2, we do not take into account the 
  earthquakes which eventually take place in the region $A$ and $B$ 
  of the two profiles, until $A$ and $B$ have a non-zero overlap.

Rule III is a consequence of the proportionality between the energy
 released during an earthquake and its seismic momentum $M_0$,
which, on its turn, is proportional to the average displacement
of a fault during an earthquake. It is obviously possible to consider
more sophisticated schemes and the work along these lines is already
 in progress.
  
  With these rules, the motion of the two profiles simulate the
  slipping of the two walls of a single fault. The points of collision are 
  the points of the fault where the morphology prevents the free 
  slip: these are the points where there is an accumulation of stress 
  and, consequently, a raise of pressure. 
  When the local pressure exceeds a certain threshold, it happens
  a breaking, an earthquake, which allows to relax the stress and 
  redistribute the energy, previously accumulated, all around.
 We assumed that the region between the two sliding profiles of the
 fault is empty or filled by a granular medium, coherently to the
observation that the fault gauge is a zone of fractured rocks. 
According to the paper by Herrmann et al. \cite{hmb} one could think
to this granular medium as composed by roller bearings between the
two surfaces. The existence of large region between the two rough 
surfaces could then be related to the so-called {\em seismic gap},
namely an extended area two tectonic plates can creep on each other
without producing either earthquakes or the amount of heat expected
from usual friction forces.
This zone slides and has no influences on the dynamics due to its relatively lower viscosity.

  For sake of simplicity, in this version of the SAM, there is no 
real breaking of the profiles as a consequence of an earthquake and  
the profiles maintain their structures after a crash.  
Thus are in the opposite perspective than SOC models. Since
  the earthquake dynamics has no effect on the structure of the profile. 
  Realistic situations could well correspond to intermediate cases, of course.

It is possible to introduce a more realistic breaking mechanism
  where there is  also a modification  of the asperity form 
  after an earthquake.  We will discuss this possibility in sect.V 
and we will show there how it is possible, in this framework, to 
reproduce the Omori's law.
   
  It is worth to stress that the SAM exhibits a strong
  non-locality since  a collision in a point $x$, at the time $t$ can 
  trigger, at later time, a subsequent event also very far away.
  One of the main advantage of the SAM consists in the 
possibility of 
  deriving various analytic results using the properties of brownian profiles.

\section{The Gutenberg-Richter law and the non-universality of the
$\beta$-exponent}

In 1956, Gutenberg and Richter \cite{GR} noticed the
dependance of earthquakes frequency
from their magnitude: the greater the magnitude, the smaller the 
frequency.
The relation between the frequency and the magnitude of 
earthquakes is:
\begin{equation}
\log_{10}{N(M>m)}=a-bm,
\label{guri}
\end{equation}
where $N(M>m)$ is the number of earthquake with a magnitude greater than $m$
while $a$ and $b$ are two empirical parameters.
The $b$ value is generally in the range $0.8<b<1.4$ depending on the 
earth region
considered and the stress level of the region itself.

Relation (\ref{guri}) is the most important statistical representation
of seismicity and the understanding of the underlying 
mechanisms is of
fundamental importance for the comprehension of earthquakes 
and their prevision.
Several studies have been achieved to understand the origin
of the universality of the Gutenberg-Richter relation but, 
despite the simplicity
of this relation, there is no understanding of the underlying mechanisms.
 The $b$ value might depend on three 
factor:
(1) the geometrical properties of the fault, (2) the physical 
properties
of the medium and (3) the stress level of the seismic region.
In this section we show that, in the framework of the SAM,
 the  $b$ value is essentially determined
by the fault geometry and in particular by its 
 fractal dimension.
The magnitude $M$ is not the only indicator of 
the earthquakes 
strength; 
another quantity used to describe the earthquake intensity is 
the seismic
moment defined by the relation:
\begin{equation}
M_0=\int_A\mu S dA,
\end{equation}
where $\mu$ is the rigidity modulus of the medium under 
consideration,
$S$ is the slip and $A$ is rupture area.
From dimensional analysis it is obvious that the energy $E$ 
released by an 
earthquake is proportional to its moment.
There is an empirical relation between the seismic moment (or 
energy) and the
magnitude:
\begin{equation}
\log_{10}{E}=cM+d,
\label{moma}
\end{equation}
where $E$ is the released energy.
From eq. (\ref{guri}) and (\ref{moma}) we 
easily obtain the energy distribution
for earthquakes:
\begin{equation}
P(E)\sim E^{-\beta-1},
\label{gr}
\end{equation}
where $P(E)$ is the probability of an earthquake releasing an 
energy $E$ and
$\beta=b/c$.

In order to describe the seismic phenomenology a model
for the fault slip has to verify equation (\ref{gr}):
we thus will study the energy distribution for the 
model defined
in the previous section (SAM).

The numerical simulations provide a clear evidence that our model
exhibits the Gutenberg-Richter law (\ref{gr}), see Fig.3. 
As we have defined in the
 previous section, the energy released
during  an earthquake is essentially given by the length of the superposition
between the fluctuations of the two self-affine profiles.
Remembering that the difference between 
two self-affine profiles is a self-affine
profile itself, we can consider only the profile given by the difference
between the upper profile and the lower profile: the energy distribution
will be simply the length distribution of the segment obtained intersecting
the difference profile with a straight line.

If we consider a fractal ensemble having a dimension $D=d-H$ embedded
in a $d$-dimensional Euclidean space, the intersection between the ensemble
and an hyperplane of dimension $d-1$ will be an ensemble of dimension 
\cite{mandel}:
$$D=(d-H)+(d-1)-d=d-H-1.$$
Therefore, the average extension of the hyper-areas given by the intersection
between a self affine hyper-surface and an hyper-plane will be:
\begin{equation}
<a>_L\sim A^{\frac{H}{d-1}},
\label{areamedia}
\end{equation}
where the subscript $L$ indicates that we are considering a portion af
hyperplane of extension $A\sim L^{d-1}$.
In virtue of the self-affine nature of the considered ensemble,
the hyper-areas distribution will be:
$$d(a)\sim a^{-\beta-1}.$$
From the last equation we can compute the mean extension of the hyperareas:
\begin{equation}
<a>_L\sim \int^{L^{d-1}}_0 a^{-\beta} da \sim L^{(d-1)(1-\beta)}
\label{areamedia2}
\end{equation}
By comparing (\ref{areamedia}) and (\ref{areamedia2}), 
one gets the relation between the exponent $\beta$
of the Gutenberg-Richter law in $d$ dimensions and the Hurst
exponent which accounts for the fractal properties of the faults:
\begin{equation}
\beta=1-\frac{H}{d-1}.
\label{B}
\end{equation}

In the three-dimensional case one has:
 \be
\beta=2-\frac{H}{2},
\ee
with $\beta\in  [\frac{1}{2},1]$.

In order to check  the (\ref{B})  we  have performed a numerical 
experiment in $d=2$.
Fig. 4 reports the results of the $\beta$-value, as a function of the
Hurst exponent, which are in good agreement with the expected 
relation $\beta=2-H$.
The dependence of $\beta$-value on the roughness
of the faults could then account for the non-universality of the 
$\beta$-value which would reflect the variability of the fractal
dimension of the fault profiles around the world. In this perspective
one could also try to relate the $\beta$-value to the age of  a given
 fault profile. By supposing that the effect of the fault slipping and of 
the earthquakes is a smoothing of the profiles, i.e.  an increase of $H$, 
one could guess that the older the fault profile, the smaller the
 $\beta$-value.

\section{Space-time distributions of epicenters} 

Let us now try to analyze the problem of the space-temporal
clustering of the earthquakes epicenters. Many authors 
\cite{cluster,Ouchi,derub}, pointed out
that the epicenter tend to cover a fractal set with a  
fractal dimension which is a highly irregular function of space
and time.
One of the most interesting feature is represented by the evidence
 that the spatial distribution of the epicenters along a linear seismogenetic structure seems to exhibit self-similar properties. Results 
of the same kind have been reported for single "transform" faults.
This could lead to the conclusion that the non-homogeneity
of the spatial distribution of the epicenters is due to some peculiar
phenomenon occurring also in a single linear fault and just partly
to the fractal distribution of the faults.

Similar properties are exhibited by the temporal distribution of events
whit a non-homogenous structure, made of periods of
 quiescence and bursts of activity. Along the same region there
could be sub-regions with non-homogeneous and also very different
behaviours.

Thanks to the simple dynamics of the SAM model it is possible to
 study, whether analytically or numerically the complex space-time
 distribution of the epicenters. Operatively the space location of an
 epicenter is defined in correspondence of the  first point of contact 
of the two colliding asperities belonging to the two profiles.

As far as spatial distribution of earthquakes is concerned, our
 simulations provide a good evidence of a spatial clustering 
of epicenters on a set with fractal dimension smaller than $1$.
In particular we obtained a value  of the fractal dimension $d_{ep}$
in the range $d_{ep} \simeq 0.8 -0.9$ for $H$ varying in
the interval [0.3,0.7] and for different lengths of the system between
$L=1000$ and $L=50000$. 

By numerical analysis of the model the values of $d_{ep}$ seem to
 decrease with increasing $H$ and seems to remain nearly 
constant with respect to variations of the system dimensions and
of $H$.
These results are not of immediate explanation; if the fault profiles
could slip for an infinite time, in fact,  each point of the inferior profile
 could be, theoretically, an epicenter 
because it, sooner or later, would be hit by an asperity of the superior
 fault profile. In this way we would have that 
$\lim_{t \rightarrow \infty} d_{ep} = \lim_{L \rightarrow
\infty} d_{ep} = 1$, 
and the set of epicenters thus becomes a compact set.
This intuitive idea turns out to be correct since the observed 
non-integer fractal dimension is a non-trivial finite-size effect.
It is possible to show analytically, for $H=0.5$, that the fractal
 dimension $d_{ep}(L)$ of the epicenter set in a fault of linear 
size $L$ is
\be
d_{ep} (L) \simeq 1- \frac{\gamma \ln \ln L}{\ln L} \,\, 
\mbox{for large $L$}.
\ee
We will sketch here the main lines of the proof referring the reader
 for further details to appendix A.

First of all it is worth to remind that, according to the definition
of the SAM, in order to obtain an infinite evolution of the system
we necessarily need two fault profiles with a length $L$,
which tends to infinity. One has first to create the two
profiles separately and then to put them in contact. This because
the average distance between the two profiles tends to increase
as $L \rightarrow \infty$.

In full generality, with reference to Fig.5, called
$<S^{\prime} (n)>=S_0^{\prime}$,  $<S^{\prime \prime}
 (n)>=S_0^{\prime\prime}$ and $h_0 =max_{j \in \lbrace
1,...,L \rbrace} \lbrace S^{\prime} (j)-S^{\prime\prime} (j) \rbrace$,
one has, from eq.(\ref{pippo}), $<S^{\prime\prime} (n)>=
S_0^{\prime\prime} +h_0$ and putting, without loss of generality,
$S_0^{\prime}= S_0^{\prime \prime}=0$,
\be
<S^{\prime \prime} (n)- S^{\prime} (n) > = h_0.
\ee

The idea we want to use is that the number of epicenters $N_{E}$,
for sufficiently large systems, will be proportional to the number of 
points of the inferior profile, $N(h,L)$, with an height $h$ between
the minimum value, $h_{min}$, of the upper fault trace, and the maximum value,
$h_{max}$, of the lower one, as shown in Fig.5: 
\be
N_{E}(L)\sim \int_{h_{min}(L)}^{h_{max}(L)} N ( h, L ) dh
\label{Nh}
\ee
where $N(h,L)$, for big values of $h$, can be written as
\be
N(h,L) \sim \sqrt{L} \exp^{- \frac{3 h^2}{4 \eta L}}
\label{pippo2}
\ee
where $3 \eta /4$ is a constant dependent on the variance of the variables $\lbrace x_i \rbrace$ used to generate the profile.
By inserting the (\ref{pippo2}) into the (\ref{Nh}) one has
\be
N_E (L) \sim \sqrt{L} \int_{h_{min}}^{h_{max}}
\exp{-\frac{3 h^2}{4 \eta L}}.
\label{pippo3}
\ee

Let us find an estimate of the $h_{min}$ and $h_{max}$ values in the
 limit $L \rightarrow \infty$ and consider the two faults to be brownian profiles ($H=1/2$) with length $L$.

In our case the variables $\lbrace X_i \rbrace$ $\lbrace Y_i \rbrace$
which compose the profiles are random variables with zero mean and
variance $\sigma^2 = 2/3$. So the variables 
$\lbrace \tilde{X_i}\rbrace= \sqrt{3/2} \lbrace X_i \rbrace$ and
$\lbrace \tilde{Y_i}\rbrace= \sqrt{3/2} \lbrace Y_i \rbrace$
will be random variables with zero mean and unitary variance.
To these variables we can apply the so-called Iterated Logarithm 
Theorem (ILT) \cite{grimmett}.
It states that,
for a partial sum $S_{k} = \sum_{i=1}^{k} w_{i}$ of identically distributed
random variables $\{ w_{i} \}$ with $<w_{i}>=0$ and variance $\sigma^{2} 
\equiv <w_{i}^{2}> = 1$, it holds:
\be
P \Bigl( \lim_{k \rightarrow \infty} \sup \frac{S_{k}}
{\sqrt{2 k \ln \ln k}} = 1 \Bigr) = 1 
\ee
where $P(A=a)$ is the probability for the variable $A$ to have the value
$a$.
For the $H=\frac{1}{2}$ case we can also write
$S_{inf}(k) = \sum_{i=1}^{k} X_{i}$
and $S_{sup}(k) = h_{0} + \sum_{i=1}^{k} Y_{i}$, where $\{ X_{i} \}$
and $\{ Y_{i} \}$ are uniformly distributed variables with zero mean,
standard deviation $\sigma^2 =2/3$ and $h_0=\max_{\forall i}
(\sum_{k=1}^i X_i-Y_i)$.

By using the ILT with profiles built with the normalized variables
$\tilde{X}_{i}$ and $\tilde{Y}_{i}$ one obtains:

\be
\begin{array}{l}
\lim_{L \rightarrow \infty} \sup_{n \in \left[ 1,L\right] }
\lbrace S^{\prime} (n)\rbrace =
\frac{2}{\sqrt{3}} \sqrt{L \ln \ln L}\\
\lim_{L \rightarrow \infty} \inf_{n \in \left[ 1,L\right] }
\lbrace S^{\prime\prime} (n)\rbrace =
-\frac{2}{\sqrt{3}} \sqrt{L \ln \ln L}.
\end{array}
\label{5.9}
\ee
One has also
\be
\inf_{n \in \left[ 1,L\right] } \lbrace S^{\prime} (n)\rbrace=
h_0 +\inf_{n \in \left[ 1,L\right] } \lbrace S^{\prime\prime} (n)\rbrace
\label{5.10}
\ee
By defining the stochastic variables 
$\lbrace Z_i = X_i - Y_i \rbrace$ it will be, by definition,
\be
h_0 = \sup_{n \in \left[ 1,L\right] } \lbrace \sum_{i=1}^{n} Z_i \rbrace.
\ee
The variables $\lbrace Z_i \rbrace$ have zero mean and variance
$\sigma^2 = 4/3$. So we can apply the ILT to the variables
$\tilde{Z_i} = \frac{\sqrt{3}}{2} Z_i$
by getting
\be
h_0= \frac{2 \sqrt{2}}{\sqrt{3}} \sqrt{\ln \ln L}.
\label{5.11}
\ee
By comparing (\ref{5.9}), (\ref{5.10}) and (\ref{5.11}) one easily gets 
the expression for $h_{min}$ and $h_{max}$:
\be
\left\{
\begin{array}{l}
h_{min}= \frac{2}{\sqrt{3}} (\sqrt{2} -1 ) \sqrt{L \ln \ln L}\\
h_{max}=\frac{2}{\sqrt{3}} \sqrt{L \ln \ln L}
\end{array}
\right.
\ee
and, inserting these expressions in (\ref{pippo3}) and making a change 
of variables,
\be
N_E (L) \sim L 
\int_{(\sqrt{2} -1 ) \sqrt{L \ln \ln L}}^{\sqrt{L \ln \ln L}}
\exp^{-\frac{t^2}{2 \eta}} dt = L I(L),
\ee
where $I(L)$ is an integral which tends to zero in the limit
$L \rightarrow \infty$. We are interested to how this integral goes
to zero.

The "average theorem" for continuous function states that it 
will be possible to find a 
$\tilde{t}= \gamma (L) \cdot \sqrt{2 \ln \ln L}$, with
$\gamma (L) \in ] \sqrt{2}-1 , 1 [$, in such a way that
\be
 N_{E}(L) \sim L \cdot e^{- \frac{\tilde{t}^{2}}
{2 \eta}} \cdot \Delta t  \sim 
\frac{L}{(\ln L)^{\tilde{\gamma}^2 (L) /\eta}}
\ee
where $\Delta t $ is the integration interval and 
$\tilde{\gamma} (L)$ is the limit value of $\gamma(L)$ and we
have neglected all the terms diverging slower than the 
logarithm.

Using the mass-length definition of fractal dimension, 
\be
d_{ep} = \lim_{L\rightarrow \infty} \ln N_{E}(L)/ \ln L
\ee, 
we obtain the relation 
\be
 d_{ep} \simeq 1 - \eta^* \frac{\ln \ln L}{\ln L} + 
O \Bigl( \frac{\ln \ln \ln L}{\ln L} \Bigr), 
\label{pippo4}
\ee
where $\eta^*$ is the mean value of $\tilde{\gamma}^2 (L) /\eta (L)$.
  This implies that, according to what we had forewarned,
 $\lim_{L\rightarrow \infty} d_{ep} = 1$, and, thus, that the 
fractal nature of the spatial distribution of epicenters is due to 
the faults finite size.
The asymptotic value $d_{ep}=1$ is reached very slowly at increasing
 $L$ and it cannot be detected but by means of huge simulations.
We have checked the validity of (\ref{pippo4}) for profiles with
a linear size $L$ varying in the range $10^2 - 10^6$. Work is in
 progress to extend our results to the case of a generic roughness
index $H$.

Let us now discuss the temporal correlations
of earthquakes and in particular problem of the
$1/f$ noise.
 A system is said to exhibit $1/f$ noise when its power spectrum  scales as 
\be
S(f)\sim\frac{1}{f^\alpha}, 
\label{noisee}
\ee
with $\alpha$ smaller than 2. 
The interest on $1/f$ noise lies in its ubiquity in nature. $1/f$
noise has been detected in systems as divers as resistors, the 
hourglass, the flow of the rivers or of the cars in a traffic system.
Despite much work has been devoted to this topic it is still
lacking a general theory that explains the widespread occurrence of $1/f$ noise.

The fact that the power spectrum is connected to the autocorrelation
 function by the Wiener and Khintchin theorem leads some authors to
 the idea that the presence of the $1/f$ noise indicates the presence of
 self-similarity in the distribution of correlation times.

The autocorrelation function is usually defined as
\be
C(t) = \frac{<E(t+t_0)E(t_0)>}{<E(t_0)>^{2}}-1,
\ee
where $E(t)$ in our case represents the energy released by an 
earthquake occurred at the time $t$ and the averages are taken 
over the distribution of times $t_0$.
If the energy presents a power-law distribution with an exponent
greater than $-2$, as in our case, the average $<E(t)>$ will depend
on its maximum value and then on the system dimension.
We would have, in this way, a non consistent procedure to calculate
 the autocorrelation function. In order to overcome his difficulty one
can use an alternative definition of the autocorrelation function which 
is independent of the scale of the system. If we define it as
\be
C(t)= <E(t+t_0) E(t_0)>
\label{autocorr}
\ee
it is possible to show \cite{sylos} that the power spectrum $S(f)$
is linked to the Fourier transformation of $C(t)$, $<|E_{f}|^2>$, by 
the relation
\be
<|E_{f}|^2>=S(f)+1/N
\ee
where $N$ represents the dimension of the system. We have also
\be
S(f)= \sum_{f^{\prime}}<|E_{f^{\prime}}|^2>|W(f-f^{\prime})|^2.
\ee
where the function $W(f)$ takes into account the finite dimension
of the system and tends to the delta function $\delta(0)$ for an infinite system. If $N$ is big enough one has:
\be
S(f)\simeq <|E_f|^2 >
\ee
and one can study the power spectrum simply analyzing the Fourier
transform of the autocorrelation function (\ref{autocorr}).

In our numerical simulation we have studied this function and the
 results, shown in Fig.6, gave 
\be
S(f)\sim f^{-\alpha}
\ee
with $\alpha \simeq 1.2$.
This means that our model exhibits $1/f$ noise, i.e. it does not exist
 a maximum autocorrelation time and a seismic event may be
 influenced by another one very distant in time.

\section{The Generalized SAM model}  

Up to now we have studied a version of the SAM model 
corresponding to the limit of infinite rigidity of the faults. The 
fault profiles are not modified by the seismic activity and one 
studies the statistics of earthquakes in the hypothesis that 
there is a complete time-scale separation between the seismic 
activity and the rearrangement of the earth crust. The latter 
would develop in very long times with respect to the scale of 
human records and this would justify the assumption. 

We have shown how this model exhibits a good interpretation 
of the seismic phenomenology in a global sense: Gutenberg-
Richter law, epicenter clustering. What is lacking is the 
description of what happens locally, i.e. as a consequence of a 
single event, both from the temporal point of view or from the 
spatial one. In particular it is not possible to obtain in such a 
scheme the Omori's law for the distribution of 
aftershocks. These events are related to the 
situation in the neighborhood of the main shock 
epicenter after the occurrence of the main shock. They are 
ruled by the following empiric relation \cite{omori}

\be
N(t) \sim \frac{1}{(t+c)^\alpha},
\label{omori}
\ee
where $N(t)$ indicates the number of earthquakes occurred at 
the time $t$ after the main shock, $c$ is a constant and 
$\alpha$ is an exponent whose value ranges in the interval 
$[1.0 \div1.4]$. For enough long times $t$ one usually
supposes $t >> c$ and the functional form of $N(t)$ is given by a pure
 power-law $N(t) \sim t^{-\alpha}$.

In this section we improve the model in order to include the 
rearrangement of the earth crust as a consequence of the 
occurrence of an earthquake. With this modification it is 
possible to describe the local phenomenology of seismicity and 
in particular to reproduce the 
Omori's law.

The model is modified by considering the asperity 
breaking in the collisions. When two asperities collide a 
fracturing process starts in the smallest asperity (that one with the 
smallest section at the level of the epicenter). The fracture 
propagates inside the fault until it crosses again the fault 
profile. At this point the fracture stops and the resulting configuration
represents the new fault profile in the region interested by the 
earthquake. The magnitude of the earthquake is assumed to be 
proportional to the linear extension of the fracture.

In Fig.7 it is shown an example of fracturing 
process during an earthquake. The shadowed region is removed 
from the fault profile. The statistical properties of the fracture 
are supposed to be identical to the ones of the entire fault 
profile. This means that one has to consider a self-affine 
profile with the same Hurst exponent of the original fault.

In our simulations, we considered, for sake of simplicity, the 
case of a Brownian profile with $H=0.5$.
Let us note that an earthquake in a certain point can trigger 
several other earthquakes, with smaller magnitude, which 
occur in the same region or in a very close region. In order to 
investigate the statistics of the aftershocks we identified all the 
aftershock occurring after a certain mainshock in the rupture region.
A mainshock is defined as an earthquake above of a certain 
magnitude (in our simulations an earthquakes involving at 
least $100$
sites). Starting from this event one counts, as a function of the 
time elapsed from the mainshock, the number of earthquakes, 
with a magnitude smaller than that of the mainshock, occurring 
in the same region. One stops the counting when an earthquake 
with a magnitude greater or equal to the mainshock occurs.

We have studied the behaviour of the cumulative distribution of 
aftershocks, i.e. the number $N_{cum}(t<T)$ of earthquakes 
occurring before $T$ time steps after the mainshock. By averaging 
over many realizations (of the order of $10^2$) we have obtained the
 curve reported in Fig.8 that exhibits the Omori scaling
 law (\ref{omori}).
For values of $t$  large enough  one has the power law:
\be
N_{cum}(t<T) \sim T^{1-\alpha}.
\label{omori2}
\ee
with the exponent $\alpha \simeq 0.37$. The 
numerical value of the exponent $\alpha$ is not in good
agreement with real values.
However, we have just considered the case of a one-
dimensional profile embedded in a two-dimensional space and the
model considers only one isolated faults, thus neglecting the effects
of interaction among different faults. It 
would be interesting to study what happens considering the case 
of a two-dimensional surface too.

This generalized SAM recalls the work
of Herrmann et al. \cite{hmb} about the space-filling bearing. The
analogy lies in the fact that one could think the interspace between
the two fault planes as filled by a granular medium which is also 
composed by the broken asperities of the fault. The link is made closer
 by the fact that in our case the distribution of areas of the asperities
 broken follows a power-law
\be
P(A_{asp}) \sim A_{asp}^{-\alpha}
\label{asp}
\ee
with an exponent $\alpha$ which could be related analytically
to the roughness exponent by the relation
\be
\alpha = \frac{2}{1+H}.
\ee
Relation (\ref{asp}) is obtained by supposing that the area of the
 broken asperities scales with its linear extension $l$ as 
$A_{asp} \sim l^{1+H}$ by a standard variable change.
Work in this direction is still in progress and we refer the reader
 to a forthcoming paper \cite{mazzella}.

It is obviously possible to consider more realistic generalizations of the 
breaking mechanism, in which the application of the pression in a 
certain point causes the breaking in a different point, mimicking, in
 this way, the effect of the stress redistribution in the medium.
This situation is, on its turn, a simplification with respect 
to the ideal case in which one has to calculate, at each time step, 
the new stress
field in the whole medium as a consequence of the changed pression
conditions.

\section{Conclusions and perspectives} 

In summary, we have proposed a model of earthquakes where 
  the critical behavior is generated by a pre-existent fractal geometry
  of the fault. The statistics of earthquakes is thus related to 
  the roughness of the fault via the scaling relation (2)
  between critical indices.  
  This result suggests that the younger the fault system, 
  the larger the $b$ exponent is, since one expects that the roughness 
  of a fault decreases in geological times.
  Note that in this case, the exponent $b$ is non-universal. 
  Another major result is that the fractal distribution of the epicenters
  could be a finite size effect very difficult to be detected from 
  data analysis. In our case our results provides a possible explanation for the
  highly irregular and non random distribution of epicenters that is observed
  experimentally. Least but not last, the accumulation of pressure
  is at the very origin of large seismic events in the SAM.
  The presence of such an effect could be tested also in real 
  situations e.g. by piezo-electric measurements.

Moreover we introduced a generalization of the SAM which includes 
the effect of the breaking of the asperities in contact during an 
earthquake. This makes the model much more realistic and allows 
for the interplay between earthquakes and structural properties of the 
faults. This version of the model exhibits a non-trivial distribution
of aftershocks which follows the Omori's law.

\newpage

\appendix{\large \bf Appendix A: Statistics of fractional Brownian
motions}

\def\theequation{A. \arabic{equation}}
\setcounter{equation}{0}

In this appendix we review the main properties of the so-called 
fractional Brownian motions (fBm for short) which represent a
 generalization of the Brownian motion \cite{manwal,mandel,voss}.

A fBm $F_H (t)$ is defined as a monodrome function of one variable
 $t$, such that its increment $\Delta F_H (\Delta t)= F_H (t+\Delta t) -
F_H (t)$ has a gaussian distribution with variance
\be
\sigma^2 = < {\Delta F_H}^2 (\Delta t) > \sim {\Delta t}^{2H},
\ee
where the brackets indicate the average over many realizations of
 $F_H (t)$. The parameter $H$ is the so-called Hurst exponent and
 takes values between $0$ and $1$. The main properties of those 
functions can be summarized as follows:

\begin{itemize}

\item[1)] they are stationary, i.e. the average square increment
 depends only on the increment of the argument $t$ and all the
values of this argument are statistically equivalent;
\item[2)] they are continuous functions  but nowhere differentiable;
\item[3)] they are self-affine curves, i.e. if the time scale is rescaled
by a factor $r$, the corresponding increment $\Delta F_H (t)$ is
 rescaled by a factor $r^H$:
\be
< {\Delta F_H}^2 (r \Delta t) > \sim r^{2H} < {\Delta F_H}^2 (\Delta t) > .
\ee

\end{itemize}

THe fBm are self-affine curves which present a box-covering
 dimension  equal to $D_F=d-H$. Let us consider, for sake of simplicity,
the case $d=2$ and suppose that $F_H (t)$ is defined in a time interval 
$\Delta t=1$ with a vertical extension $\Delta F_H (t) =1$. If one
rescales the time by a factor $r <1$, then, by virtue of the self-affinity,
$F_H (t)$ will be rescaled by a factor $r^H$. Thus, in order to cover a
 section of curve extending in the interval $\Delta t =r$ one needs
$\Delta F_H / \Delta t= r^{H-1}$ boxes of linear dimension$r$ and
for the entire profile one will need $r^{H-1} /r$ boxes.
So recalling the definition of box-covering dimension
\be
D_F = \lim_{r \rightarrow 0} \frac{\log N(r)}{\log 1/r}
\ee
one has
\be
D_H = \lim_{r \rightarrow 0}  \frac{\log r^{H-2}}{\log 1/r}= 2-H.
\ee
In the general $d$-dimensional case one can define the brownian
hyper-surface as a function of $n=d-1$ variables $X_i \,\, i=1,...,n$
such that
\be
<{\Delta F_H}^2 (\Delta r) > \sim {\Delta r}^{2H},
\ee
with ${\Delta r}^2= {\Delta X_1}^2+...+{\Delta X_n}^2$.

The box-covering fractal dimension is then defined as 
\be
D_F= n+1-H= d-H.
\ee

A last word about the intersection of a fBm with a line parallel
to the temporal axis (fractal dimension $D_1$) and lying in the same 
plane of the brownian  profile (fractal dimension $D_2$). 
In this case, by using the law of additivity of the codimension, 
\be
D_0 = D_1 + D_2 -d,
\ee
where $D_0$ is the fractal dimension of the intersection set, 
the zeroset.
In our case one has $D_1=d-1$ and $D_2 = d-H$ and then
\be
D_0= d-1-H.
\ee
The set of zeroes of a Brownian profile in $d=2$ with a generic
vale of the Hurst exponent $H$ is then a set of points whose 
fractal dimension is $D_0 =1-H$

\newpage

\appendix{\large \bf Appendix B}

\def\theequation{B. \arabic{equation}}
\setcounter{equation}{0}

In this appendix we calculate the number of points
that, in a Brownian profile, lie at a certain height
$h$. from the general properties of the brownian profile one knows
that if $h=0$, this number is proportional to $\sqrt{L}$ where $L$
is the length of the profile. Moreover, as a consequence of the spatial
homogeneity of the random walk one has
\be
P(S_{n+m}=0 \vert S_n =0 ) = P(S_{n+m}=h \vert S_n =h )  
\ee
where $P(a \vert b)$ is the conditional probability that, given a 
certain event $b$, the event $a$ occurs.
The number of points at the height $h$ will be proportional
to $\sqrt{L-t}$ where $t$ is the first passage time at the height $h$.
The first passage time distribution for an height $h$ is known
\cite{grimmett} to be
\be
f_h (t) = \frac{\vert h \vert}{\sqrt{2 \pi t^3}} 
\exp^{- \frac{h^2}{2 t}}.
\ee
 One then has that the number of points at the height $h$ is
\be
N(h) \sim \int_{0}^{L} \sqrt{L-t} f_h (t) dt.
\label{ciccia}
\ee
Eq.(\ref{ciccia}) is a very complicate expression and we limit
 ourselves to consider what happens just in the range $h \ge \sqrt{L}$.
For the average theorem it will exist a value $t^*$ such that
\be
N(h) \sim \sqrt{L-t^*} f_h (t^*) L.
\ee
We are interested in the case of $h \sim \sqrt{L}$ and we can then
 suppose $t^* \sim \eta L$. One obtains
\be
N(h) \sim \sqrt{L} \exp^{-\frac{h^2}{2 \eta L}}
\ee
that we used in sect.4.

\vspace{1.0cm}

{\bf \large Acknowledgements}

It is a pleasure to thank V. De Rubeis and P. Tosi with which
part of this work has been carried out. We thank E. Caglioti and
G. Mantica for useful suggestions.

\newpage

\section*{FIGURES}

\begin{figure}[htb]

\centerline{
        \epsfxsize=12.0cm
        \epsfbox{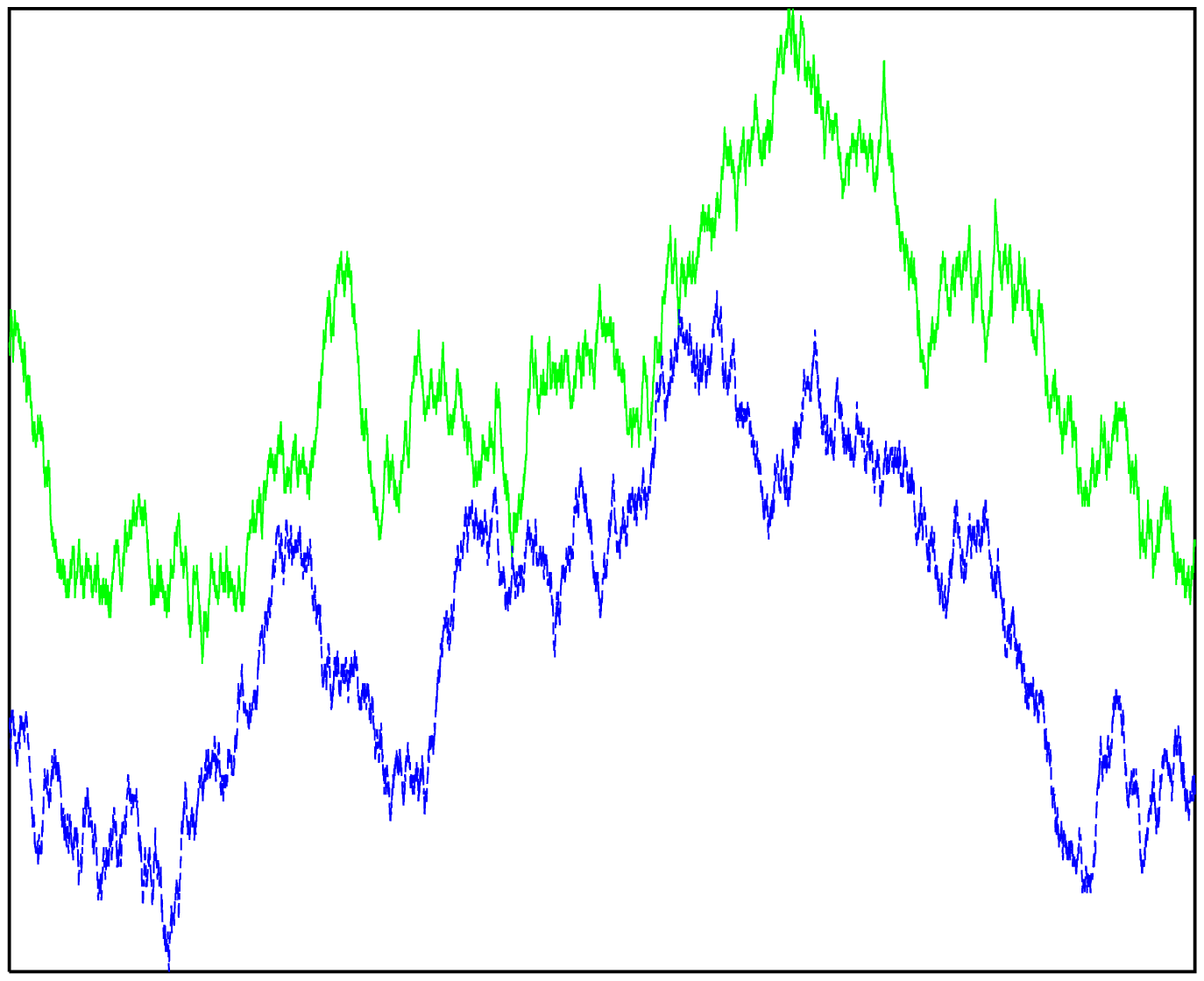}
        \vspace*{0.5cm}
}
\caption{Fault planes realized by two Brownian profiles
put in contact in one point.}
\label{Fig.1}
\end{figure}

\begin{figure}[htb]

\centerline{
        \epsfxsize=12.0cm
        \epsfbox{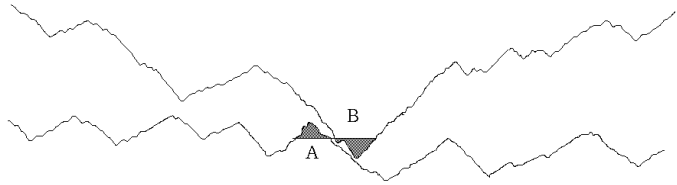}
        \vspace*{0.5cm}
}
\caption{Sketch for the definition of the energy released
 during an earthquake. It is assumed proportional to the breaking area
(the $(d-1)$-dimensional sets $A$ and $B$) between the two asperities:
$E \propto A+B$.}
\label{Fig.2}
\end{figure}

\begin{figure}[htb]

\centerline{
        \epsfxsize=12.0cm
        \epsfbox{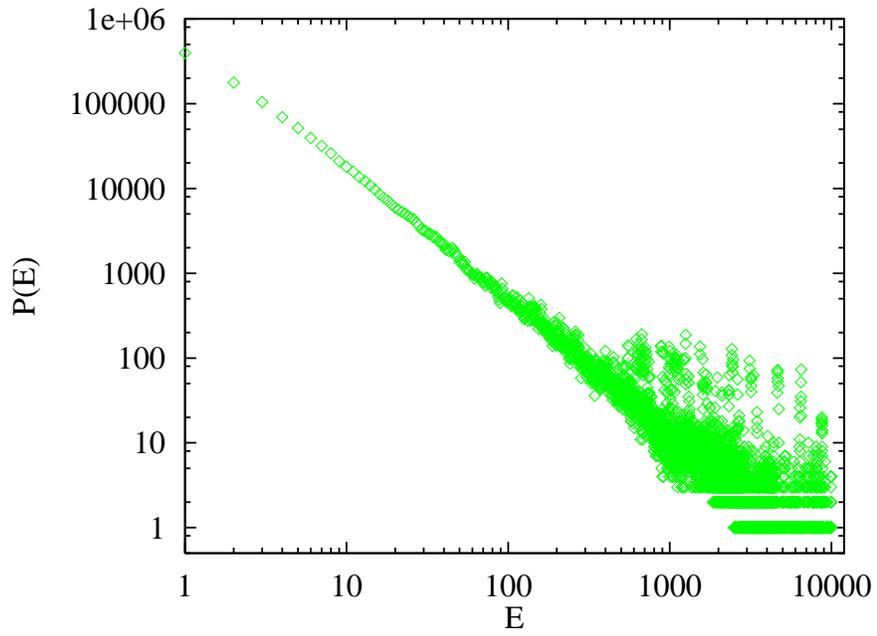}
        \vspace*{0.5cm}
}
\caption{Probability density of the earthquakes releasing 
an energy $E$ vs. $E$ for roughness index $H=1/2$.}
\label{Fig.3}
\end{figure}

\begin{figure}[htb]

\centerline{
        \epsfxsize=12.0cm
        \epsfbox{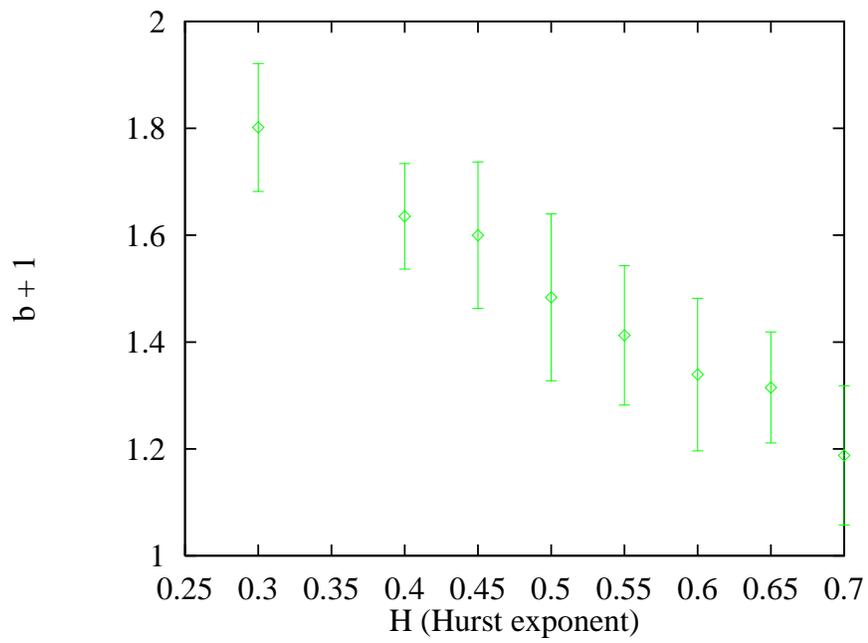}
        \vspace*{0.5cm}
}
\caption{Exponent $\beta +1$ vs. the Hurst exponent $H$
for the SAM in $d=2$.}
\label{Fig.4}
\end{figure}

\begin{figure}[htb]

\centerline{
        \epsfxsize=12.0cm
        \epsfbox{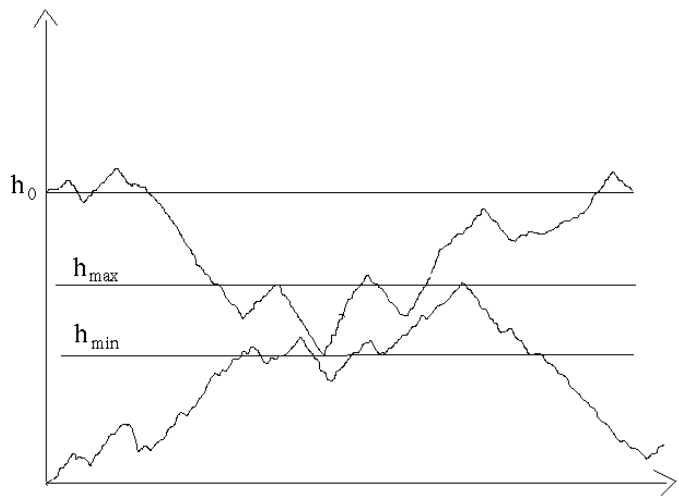}
        \vspace*{0.5cm}
}
\caption{Scheme illustrating the region of heights  
$[ h_{min}, h_{max}]$ in which it is possible the occurrence 
of collisions between asperities. $h_0$ is a function of the
length $L$ of the profiles and indicates their average
distance.}
\label{Fig.5}
\end{figure}

\begin{figure}[htb]

\centerline{
        \epsfxsize=12.0cm
        \epsfbox{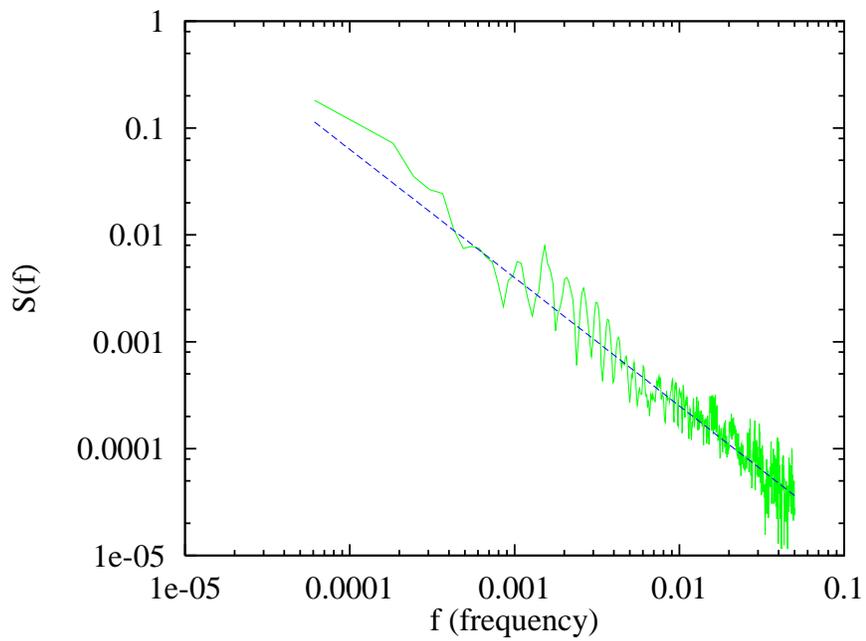}
        \vspace*{0.5cm}
}
\caption{Power spectrum (solid line) for the temporal sequence 
of earthquakes in the SAM model. It shows a $1/f$ behaviour 
with an exponent $\alpha \simeq 1.2$ corresponding to the slope 
of the dashed line.}
\label{Fig.6}
\end{figure}

\begin{figure}[htb]

\centerline{
        \epsfxsize=12.0cm
        \epsfbox{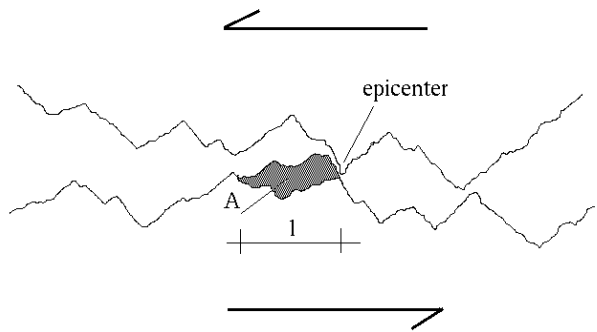}
        \vspace*{0.5cm}
}
\caption{Scheme illustrating the mechanism for the breaking
of the asperities in the generalized SAM model. When there is 
a collision between two asperities the weaker is broken. 
The shadowed region defines the broken area and the new 
profile after the collision.}
\label{Fig.7}
\end{figure}

\begin{figure}[htb]

\centerline{
        \epsfxsize=12.0cm
        \epsfbox{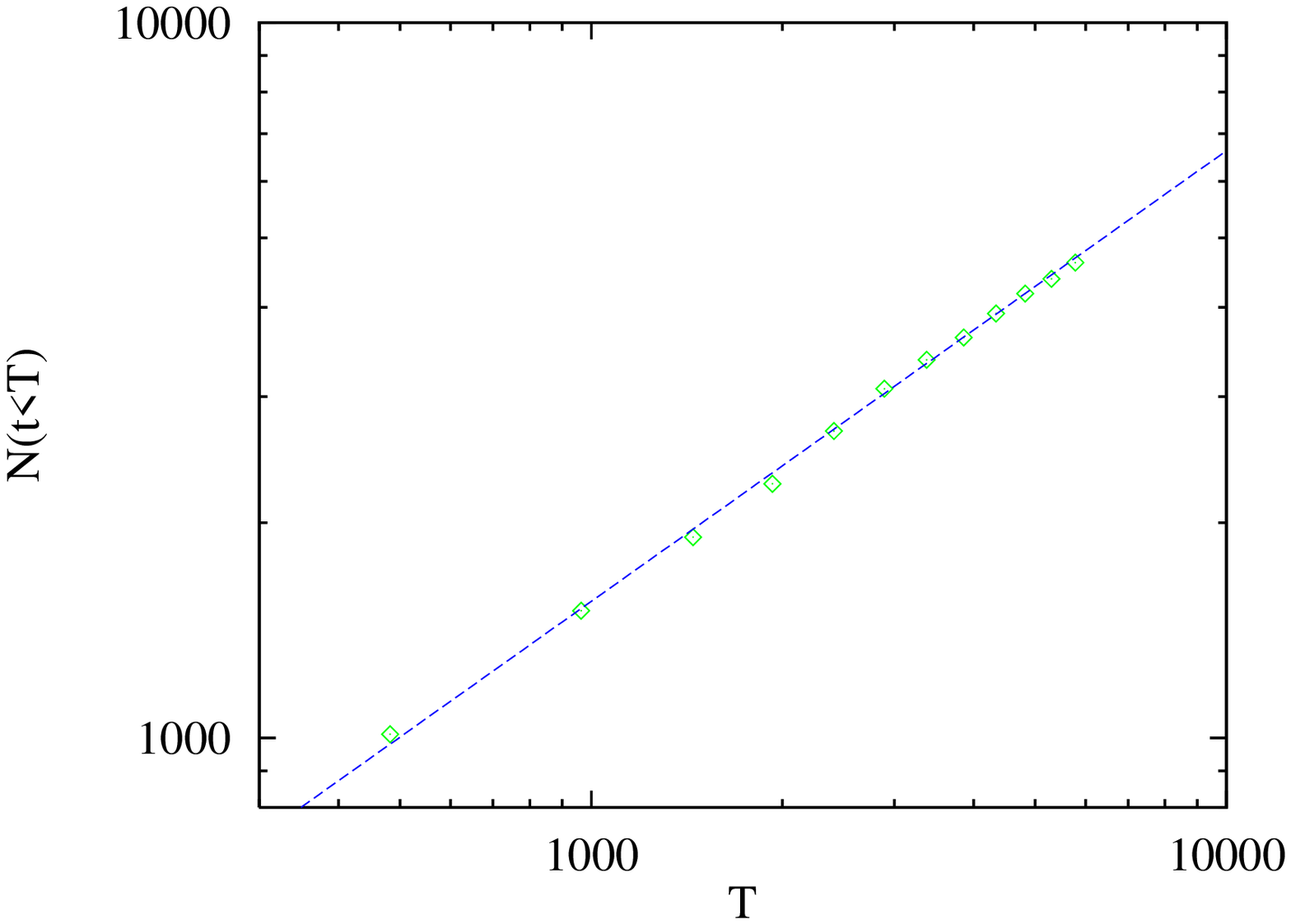}
        \vspace*{0.5cm}
}
\caption{Cumulative distribution for the aftershocks: $N(t < T)$
is the number of aftershocks, events causally connected to the
mainshock, occurred up to the time $T$, elapsed from the
mainshock time.}
\label{Fig.8}
\end{figure}


\newpage
 
 \begin{thebibliography}{99}
\bibitem{GR} Gutenberg B. and Richter C.F.  1956 {\em Ann. Geophys.} 
 {\bf 9}, 1. 

\bibitem{omori} Omori F. 1894,  {\em Rep. Earth. Inv. Comm.},  
{\bf 2} , 103.

\bibitem{cluster} Kagan Y.Y. and Knopoff L. 1980, {\em Geophys. 
J. R. Astron. Soc.},  {\bf 62}, 303.


\bibitem{burkno} Burridge R. and Knopoff L. 1967, {\em 
Bull. Seismol. Soc. Am.}  {\bf 57}, 341.

\bibitem{carlson} Carlson J.M. and Langer J.S. 1989, {\em Phys. Rev.
 Lett.}  {\bf 62},  2632; {\em Phys. Rev. A}  {\bf 40}, 6470.

\bibitem{paladin} Crisanti A., Jensen M.H., Vulpiani A. and Paladin G.,
 1993  {\em Phys. Rev. A} {\bf 46} R7363.

\bibitem{btw} Bak P., Tang C. and Wiesenfeld K. 1987, 
{\em Phys. Rev. Lett.} {\bf 59}, 381; 1988 {\em Phys. Rev. A} {\bf 38}, 364.

\bibitem{olami} Olami Z., Feder H. J. S. and Christensen 1992, 
{\em Phys. Rev. Lett.}  {\bf 68} 1244; Christensen K. and Olami Z. 1992, 
{\em J. Geophys. Res.} {\bf 97}, 8729.

\bibitem{baktang} Bak P. and Tang C. 1989, {\em J. Geophys. Res.}  
{\bf 94}, 15635.
  
\bibitem{ito} Ito K. and Matsuzaki M. 1990, {\em J. Geophys. Res.} 
{\bf 95}, 6853.

\bibitem{prl} V.De Rubeis, R. Hallgass, V. Loreto, G.Paladin,
 L.Pietronero and P. Tosi: {\em Phys. Rev. Lett.} {\bf 76}, 2599 1996.

\bibitem{brown} Brown S.R. and Scholz C.H. 1985, {\em  J. Geophys. Res.}, 
{\bf 90}, 12575; Wu R.S. and Aki K. 1985, {\em PAGEOPH}  {\bf 123}, 805.

\bibitem{power} Power W., Tullis T., Brown S., Boitnott G. and Scholz
 C.H. 1987, {\em Geophys. Res. Lett.}  {\bf 14}, 29.

\bibitem{huangturcotte} J. Huang and D.L. Turcotte, {\em Earth Planet.
Sci. Lett.}, {\bf 91} 223 (1988).


\bibitem{Ouchi} T. Ouchi and T. Uekawa, {\em Phys. Earth Planet.
Inter.}, {\bf 44} 211 (1986).


\bibitem{derub} V. De Rubeis, P. Dimitriu, E. Papadimitriu and P. Tosi,
{\em Geophys. Res. Lett.}, {\bf 20} 1911 (1993).

\bibitem{mandel} Mandelbrot B. 1983, {\em The fractal geometry
 ofÊnature}, Freeman and Co., New York, pp 256-258.

\bibitem{turcotte} Turcotte D. L. 1992, {\em Fractals and chaos in
 geology and  geophysics}, Cambridge Un. Press, Cambridge.

\bibitem{smalley}
   Smalley R.F. Jr., Chatelain J.L., Turcotte D.L. and Pr\'evot R.: 
   1987 {\em Bull. Seis. Soc. Am.}  {\bf 77}, 1378; Hirata T.: 1989 {\em
J. Geophys. Res.} {\bf 94}, 7507; Kagan Y.Y. and Jackson D.D.: 1991,
  {\em Geophys. J. Int.} {\bf 104}, 117; Main I. G. 1992, {\em Geophys.
 J. Int.}  {\bf 111}, 531; De Rubeis V., Dimitriu P., Papadimitriou E. and
 Tosi P.  1993, {\em Geophys. Res. Lett.} {\bf 20}, 1911.

\bibitem{hmb} H.J. Herrmann, G. Mantica and D. Bessis, {\em Phys.
Rev. Lett.}, {\bf 65}, 3223 (1990).

\bibitem{grimmett} Grimmett G. R., Stirzaker D. R.1992, {\em
 Probability and Random Processes second edition}, Oxford Science
 Publications, New York.

\bibitem{sylos} L. Amendola and F. Sylos: {\em Power Spectrum 
for Fractal Distributions}, in print on 
{\em Astrophys Lett \& Comm} 1996.

\bibitem{mazzella} O. Mazzella,  R. Hallgass, V. Loreto, G. Paladin and 
L. Pietronero, {\em The Omori's law  in the framework of the
 Self-affine Asperity Model}, preprint (1996).

\bibitem{manwal} B.B. Mandelbrot and J.W. Van Ness, {\em SIAM Rev.}, 
{\bf10} 422 (1968).

\bibitem{voss} R.F. Voss, {\em Physica D} {\bf 38}, 362 (1989).

 \end{thebibliography}
 \end{document}